\documentclass[reprint, amsmath,amssymb,aps,]{revtex4-2}

\usepackage{graphicx}
\usepackage{dcolumn}
\usepackage{bm}
\usepackage{graphicx}
\usepackage[utf8]{inputenc}
\usepackage[export]{adjustbox}
\usepackage{wrapfig}
\usepackage{amsmath}

\begin{document}

\preprint{APS/123-QED}

\title{Complete Population Transfer of Maximally Entangled States in ${\mathrm{2}}^{\mathrm{2}N}$-level Systems via Pythagorean Triples Coupling}

\author{Muhammad Erew}
\email{erew@tauex.tau.ac.il}
\author{Moshe Goldstein}
\email{mgoldstein@tauex.tau.ac.il}
\homepage{\\http://www3.tau.ac.il/mgoldstein/index.php}
\author{Haim Suchowski}
\email{haimsu@post.tau.ac.il}
\homepage{\\https://m.tau.ac.il/~haimsu/Home.html}
\affiliation{
Raymond and Beverly Sackler School of Physics and Astronomy, Tel Aviv University, Tel Aviv 6997801, Israel
}

\date{\today}

\begin{abstract}
Maximally entangled states play a central role in quantum information processing. Despite much progress throughout the years, robust protocols for manipulations of such states in many-level systems are still scarce. Here we present a control scheme that allow efficient manipulation of complete population transfer between two maximally entangled states. Exploiting the self-duality of $\mathrm{SU}\left(2\right)$, we present in this work a family of ${\mathrm{2}}^{\mathrm{2}N}$-level systems with couplings related to Pythagorean triples that make a complete population transfer from one state to another (orthogonal) state, using very few couplings and generators. We relate our method to the recently-developed retrograde-canon scheme and derive a more general complete transfer recipe. We also discuss the cases of $\left(2n\right)^2$-level systems, $\left(2n+1\right)^2$-level systems and other unitary groups.
\end{abstract}

\maketitle

\section{Introduction}
Quantum coherent control attracts a great experimental and theoretical interest in current days, especially toward multi-state quantum systems \cite{quantum_control_1,complexity, security, quantum_information_1,quantum_control_2}. Complete population transfer (CPT) plays an indispensable role in this effort \cite{quantum_information_1, population_trapping_1,  spectroscopy_0, spectroscopy_2, nuclear_magnetic_resonance_1, quantum_computing_0,quantum_computing_2, chemical_dynamics_1}, making it is highly desirable to introduce novel efficient methods and models for building such transfer themes. For general time-dependent coupled dynamical equations, it is not easy to find solutions analytically. Even for a simple two level system there is a limited number of known time-dependent Hamiltonians that can be solved analytically and give a CPT \cite{atomic_exciatation_1,two_level_2}. Of special importance in this respect are maximally entangled states, which play a central role in quantum information processing. Despite much progress throughout the years, robust protocols for manipulations of such states in many-level systems are still scarce. Thus, controlled manipulation between such states and specifically CPT are naturally desirable.

Theoretical methods have been found and developed for complete controllability of systems \cite{theoretical_1,theoretical_2}. However, these methods are nonconstructive and do not help in a concrete system. Due to the difficulty of synthesis and analysis of CPT schemes increases in multi-state systems, multi-state control problems are usually reduced to two-state ones \cite{reducing_2,elimination, restrictions_1, inaccuracies, N-level_1}. Several approaches have been also proposed for CPTs in multi-level systems \cite{3-level-CPT_1,3-level-CPT_2,CPT_Degenerate_1,CPT_multi-level_1,CPT_multi-level_2,CPT_multi-level_3}. Recently, the dynamics of four level atomic system has been explored from a geometrical point of view, revealing that one can obtain CPTs in the lab frame if and only if some constraints on the couplings are obeyed \cite{pythagorean}. In the case of periodic nearest-state coupling, the requirements of CPT were found to be linked to primitive Pythagorean Triples \cite{pythagorean}. Later on, the Pythagorean coupling scheme was verified experimentally in the realm of four-level superconducting Josephson circuit \cite{application}.

In this work, using the self duality of $\mathrm{SU}\left(2\right)$, we derive a general scheme for CPTs in ${\mathrm{2}}^{\mathrm{2}N}$-level systems. We show that the basis for the CPT is composed of maximally-entangled  states. This observation is crucial for entangled-state manipulation, and we expect it to serve as a building block for future efficient quantum information processing protocols. It turns out that our novel finding is a generalization of the Pythagorean Triple coupling scheme \cite{pythagorean} to higher representation of $\mathrm{SU}\left(2\right)$, offering a new group-theoretical perspective on CPTs. We discuss the case of more general $\left(2n\right)^2$-level systems, relating our method to the recently developed Retrograde Canon scheme \cite{retrograde}. We also explain why our method does not apply to either $\left(2n+1\right)^2$-level or to higher unitary groups, but derive a new general CPT recipe for general multi-state systems. Our scheme employs a substantially-reduced number of couplings, allowing enormous simplification of its experimental realization in maximally entangled state control in various fields, including laser induced finite level systems \cite{laser_induced}, Josephson junctions \cite{application}, and waveguide arrays \cite{arrays}.

\section{The Pythagorean coupling again -- The diamond 4-level systems}
In our current derivation, we reintroduce the Pythagorean coupling found in Ref. \cite{pythagorean} from a different angle, which would allow its significant extension later on. We work with the two spin-$\frac{1}{2}$ Hamiltonians
\begin{subequations}
\begin{equation}
h^{\left(1\right)}_{2\times 2}=\Delta_1{\sigma }_z+\Omega_1{\sigma }_x  ,
\end{equation}
\begin{equation}
h^{\left(2\right)}_{2\times 2}=\Delta_2{\sigma }_z+\Omega_2{\sigma }_x  ,
\end{equation}
\label{eqn:spin_0.5_hamiltonians}
\end{subequations}
where $\Delta_1,\Omega_1,\Delta_2,\Omega_2$ are nonzero real numbers. $\Omega_1$ and $\Omega_2$ represent real Rabi frequencies, and $\Delta_1$ and $\Delta_2$ represent the detunings. We construct the $4\times 4$ Hamiltonian $H$:
\begin{equation}
    H=h^{\left(1\right)}\oplus h^{\left(2\right)}= h^{\left(1\right)}_{2\times 2}\otimes I^{\left(2\right)}_{2\times 2}+I^{\left(1\right)}_{2\times 2}\otimes h^{\left(2\right)}_{2\times 2}  ,
\label{eqn:TP_hamiltonian}
\end{equation}
and we denote this frame by the Tensor Product (TP) frame.

With proper basis change, we obtain a lab-frame picture with nearest neighbor coupling that can be realized physically by a laser-field-driven four-level atom. This could be done by the orthogonal symmetric transformation matrix $W$ composed of maximally entangled states (of which the von Neumann entropy is $\ln n$, where $n$ is the dimension of the Hilbert space \cite{quantum_information_1}, and here it is 2):
\begin{equation}
W=\frac{1}{\sqrt{2}}\left(V\left({\mathrm{\Sigma }}_0\right),V\left({\mathrm{\Sigma }}_1\right),V\left({\mathrm{\Sigma }}_2\right),V\left({\mathrm{\Sigma }}_3\right)\right)  ,
\label{eqn:W_4cross4}
\end{equation}
where the $V\left(\cdot \right)$ denotes the vectorization function described in Appendix \ref{appendix:vectorization}, and $\Sigma_0=\sigma_0,\Sigma_1=\sigma_1,\Sigma_2=i\sigma_2,\Sigma_3=\sigma_3$ ($\sigma_0$ is the $2\times2$ unit matrix, and $\sigma_1,\sigma_2,\sigma_3$ are the Pauli matrices). A very useful property of this operator is $V\left(AXB\right)=\left(B^T\otimes A\right)V\left(X\right)$, and we use it frequently in this paper.

The Hamiltonian in the two frames is:
\begin{equation}
H_{TP}=\left(\begin{array}{cccc}
V_{14} & \Omega_2 & \Omega_1 & 0 \\ 
\Omega_2 & V_{23} & 0 & \Omega_1 \\ 
\Omega_1 & 0 & -V_{23} & \Omega_2 \\ 
0 & \Omega_1 & \Omega_2 & -V_{14} \end{array}
\right)  ,
\end{equation}

\begin{equation}
H_{\mathrm{Lab}}=\left(\begin{array}{cccc}
0 & V_{12} & 0 & V_{14} \\ 
V_{12} & 0 & V_{23} & 0 \\ 
0 & V_{23} & 0 & V_{34} \\ 
V_{14} & 0 & V_{34} & 0 \end{array}
\right)  ,
\end{equation}
where $V_{12},V_{23},V_{34},V_{14}$ are defined as follows:
\begin{equation}
    \left(
    \begin{array}{cc}
        V_{12} & V_{23} \\
        V_{34} & V_{14}
    \end{array}
    \right)
    =
    \left(
    \begin{array}{cc}
        \Omega_1+\Omega_2 & \Delta_1-\Delta_2 \\
        -\Omega_1+\Omega_2 & \Delta_1+\Delta_2
    \end{array}
    \right)
\end{equation}

Now, the dynamics described by the Schr\"{o}dinger equation $\frac{\partial \psi }{\partial t}=-iH\psi$ lead to the unitary time-evolution operator (propagator) $U_{TP}=u^{\left(1\right)}\otimes u^{\left(2\right)}=e^{-ih^{\left(1\right)}t}\otimes e^{-ih^{\left(2\right)}t}$ (and $U_{\mathrm{Lab}}=WU_{TP}W$). We are interested in CPTs between basis states in the lab frame. Let $e_i$ denote the $4\times 1$ matrix (column vector) which is zero everywhere except the $i$-th component, which is 1 (in other words ${\left(e_i\right)}_{j1}={\delta }_{ij}$). Performing the calculation, if we start with ${\psi }_{\mathrm{Lab}}\left(t=0\right)=e_1\equiv{\left(1,0,0,0\right)}^T$, we find that one cannot reach either $e_2$ or $e_4$ starting from $e_1$, but can fully transfer into $e_3$ if and only if: 
\begin{equation}
\begin{split}
\Delta_1=\frac{1}{2}\frac{k\left(c-a\right)+b}{\sqrt{1+k^2}}\ \ ,&\ \ \Omega_1=\frac{1}{2}\frac{c-a-kb}{\sqrt{1+k^2}}\ \ , \\
\Delta_2=\frac{1}{2}\frac{k\left(c+a\right)-b}{\sqrt{1+k^2}}\ \ ,& \ \ \Omega_2=\frac{1}{2}\frac{c+a+kb}{\sqrt{1+k^2}} \ \ ,
\\
\tau =&\frac{\pi }{\sqrt{2c}} \ ,
\end{split}
\label{eqn:pythagorean}
\end{equation}
where $\tau$ is the CPT time, $k$ is an arbitrary real number, and $\left(a,b,c\right) = \left(\frac{p^2-q^2}{2},pq,\frac{p^2+q^2}{2}\right)$ where $p$ and $q$ are odd integers ($p>q$). We see that the triple $\left(a,b,c\right)$ has the the well-known general form of a Pythagorean triple, and it is called primitive when $p$ and $q$ are coprime, ${\mathrm{gcd} \left(p,q\right)\ }=1$. Notice that not just primitive Pythagorean triples (PPTs) give solutions, but also non primitive ones. However, Hamiltonians generated from triples that are not primitive are simply Hamiltonians that are generated from PPTs multiplied by an odd integer constant. So we can restrict ourselves just to PPTs. On the other hand, taking negative numbers, like $\left(a,b,c\right)=\left(-\frac{p^2-q^2}{2},pq,\frac{p^2+q^2}{2}\right)$ for example, suggests new inequivalent Hamiltonians, so those should be included as well.

\section{Generalizing to other representations}
The way we developed the scheme of CPT suggests a natural generalization to other representations of $\mathfrak{su}\left(2\right)$. Motivated by the fact that the Lie algebra structure is more fundamental than its representation, we investigate the CPT condition for higher dimensional representation of $\mathfrak{su}\left(2\right)$. So we consider now this Hamiltonian:

\begin{subequations}
\begin{equation}
h^{\left(1\right)}_{n\times n}=2\Delta_1J^{(n)}_3+2\Omega_1J^{(n)}_1
\end{equation}
\begin{equation}
h^{\left(2\right)}_{n\times n}=2\Delta_2J^{(n)}_3+2\Omega_2J^{(n)}_1
\end{equation}
\begin{equation}
    H=h^{\left(1\right)}_{n\times n}\otimes I^{\left(2\right)}_{n\times n}+I^{\left(1\right)}_{n\times n}\otimes h^{\left(2\right)}_{n\times n} ,
\end{equation}
\end{subequations}
and we want to get a CPT in $n$-dimensional representations. Here the matrices $\left\{J^{\left(n\right)}_1,J^{\left(n\right)}_2,J^{\left(n\right)}_3\right\}$ are the known basis of the $n$-dimensional irreducible (spin-$\frac{n-1}{2}$) representation of $\mathfrak{su}\left(2\right)$. They satisfy $\left[J^{\left(n\right)}_i,J^{\left(n\right)}_j\right]=i{\varepsilon }_{ijk}J^{\left(n\right)}_k$, with real $J^{\left(n\right)}_1$, imaginary $J^{\left(n\right)}_2$, and diagonal $J^{\left(n\right)}_3$. We mention in passing that calculating the time evolution operator of such Hamiltonians becomes much easier when one uses the Cayley–Hamilton theorem \cite{cayley_1,Cayley_2}.

\begin{figure}[t]
\centering
    \includegraphics[width=8.6cm]{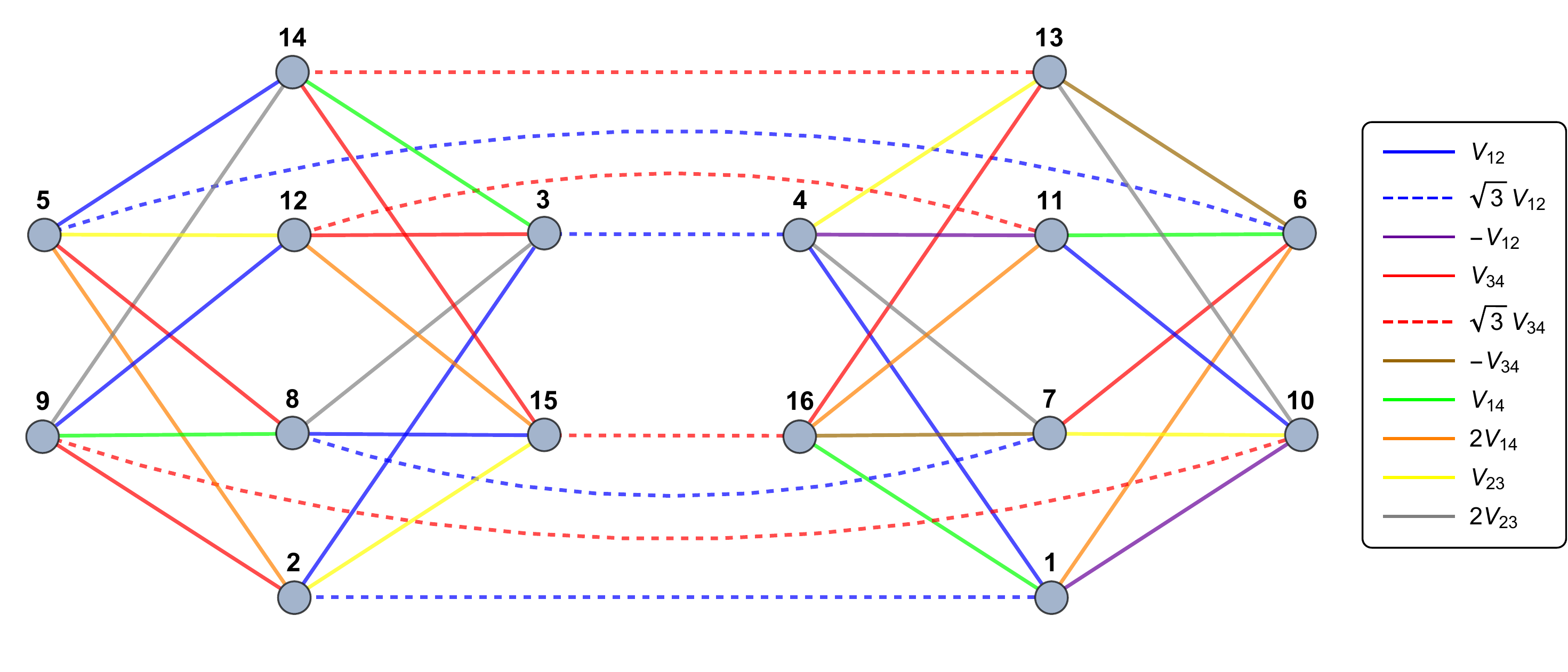}
    \caption{The couplings of the 16-level system. A CPT occurs between level 1 and level 13. One can see that the coupling structure requires much less independent couplings than a general 16-level system, which will allow a simplified experimental realization.}
    \label{fig:16_level}
\end{figure}

The main challenge in the higher-dimensional case is finding a generalized matrix $W$ which would give a lab frame Hamiltonian with a realistic structure and symmetry of its nonvanishing matrix elements (see for example Figure \ref{fig:16_level}). For $n=2^N$, this could be achieved if one constructs $W$ out of maximally-entangled states. This $W$ is composed of vectorization of tensor products of $N$ matrices from the set $\left\{{\mathrm{\Sigma }}_0,{\mathrm{\Sigma }}_{\mathrm{1}},{\mathrm{\Sigma }}_{\mathrm{2}},{\mathrm{\Sigma }}_{\mathrm{3}}\right\}$ (with a normalization factor $\frac{1}{\sqrt{2^N}}$). The resulting vectors are automatically orthogonal, making $W$ an orthogonal matrix. One can order them in a way that makes $W$ symmetric as well. We can fix $W$ by imposing the following demands: (a) the first $n$ columns do not contain negative values, (b) the first half of the diagonal contains just positive values and the second just negative values, (c) the last column contains alternating $1$s and $(-1)$s, in addition to zeros. This structure is a natural generalization of the $4 \times 4$ case presented above. In this Lab frame, and because of our convention of building the rotational matrix, we always achieve a CPT from $e_1$ to $e_{n^2-n+1}$.

To simplify the expressions, let us denote the column vector $V\left(\bigotimes^k_{i=1}\Sigma_{n_i}\right)$ by $\underline{n_1 n_2 ... n_k}$, where $n_1$ ... $n_k$ can assume the values 0, 1, 2, and 3. For example, by $\underline{0031}$ we mean $V\left(\Sigma_0 \otimes \Sigma_0 \otimes \Sigma_3\otimes \Sigma_{1}\right)$. With this notation, $W$ for the case of $N=1$ (which means that $n=2$ and we work with a 4-level system) is $W_{4\times 4}=\frac{1}{\sqrt{2}}\left(\underline{0},\underline{1},\underline{2},\underline{3}\right)$ (see Equation (\ref{eqn:W_4cross4})).

In the language of the TP frame, the CPT we are talking about is always from the state proportional to $\underline{00\dots 00}$ to the state proportional to $\underline{11\dots 112}$ ($Y_n$ always has 1s and ($\mathrm{-}$1)s alternately on the anti-diagonal, and 0s elsewhere). In the case of $N=1$ it is from $\underline{0}$ to $\underline{2}$ in the TP frame, which means it is from $e_1$ to $e_3$ in the lab frame.

As an example, when $N=2$ (this means that $n=4$ and we work with a 16-level system), one may use this orthogonal symmetric transformation to go to the Lab frame:
\begin{equation}
\begin{split}
    W_{16\times 16}=\frac{1}{2}( &\underline{00},\underline{01},\underline{10},\underline{11},\underline{31},\underline{30},\underline{21},\underline{20},\\
&\underline{23},\underline{22},\underline{33},\underline{32},\underline{12},\underline{13},\underline{02},\underline{03} )  .
\end{split}
\end{equation}
The matrix is written out explicitly in Appendix \ref{appendix:16-level_hamiltonians}.

\begin{figure*}[t]
    \centering

    \includegraphics[width=7.2cm]{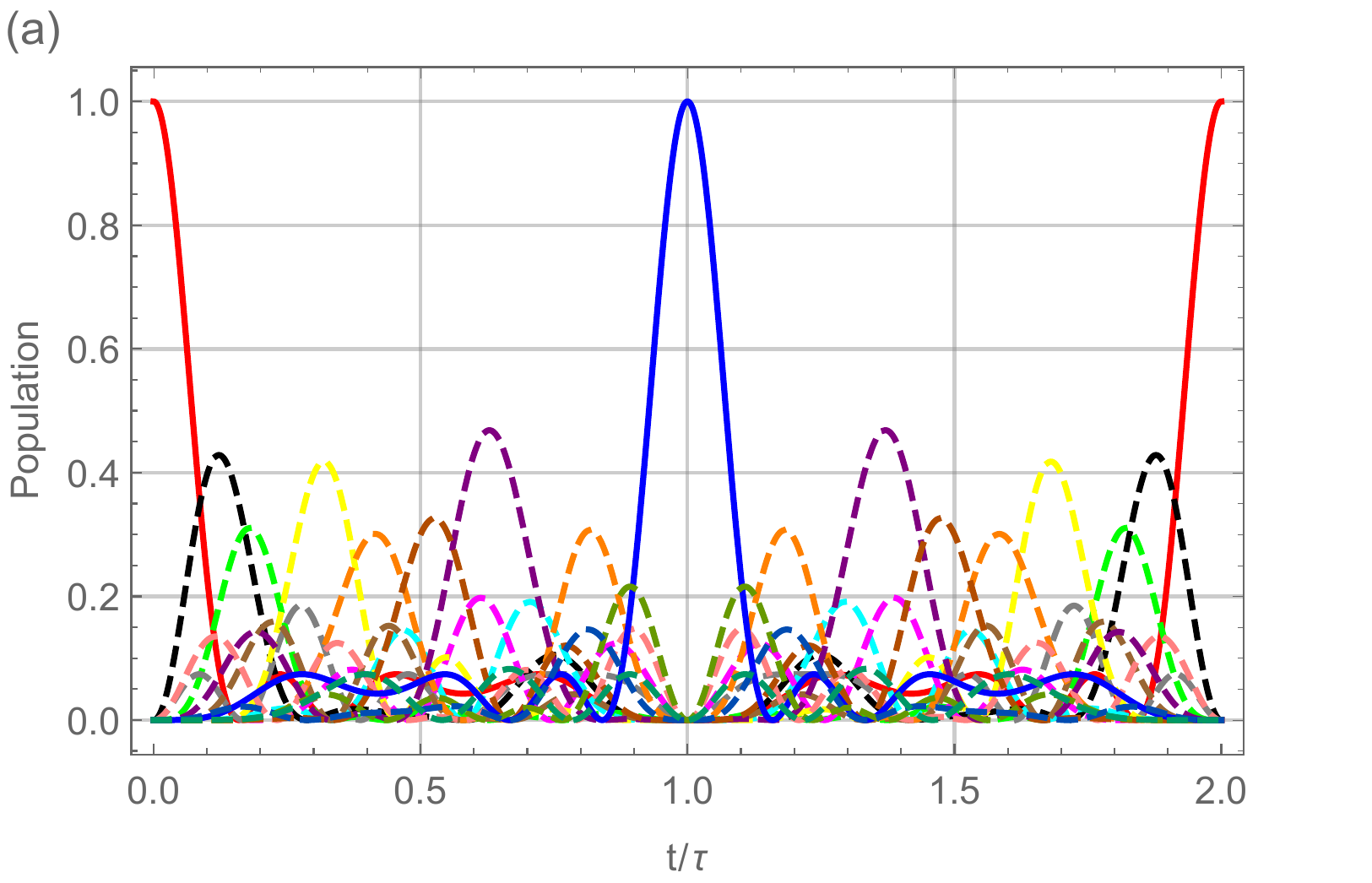}   
    \includegraphics[width=7.2cm]{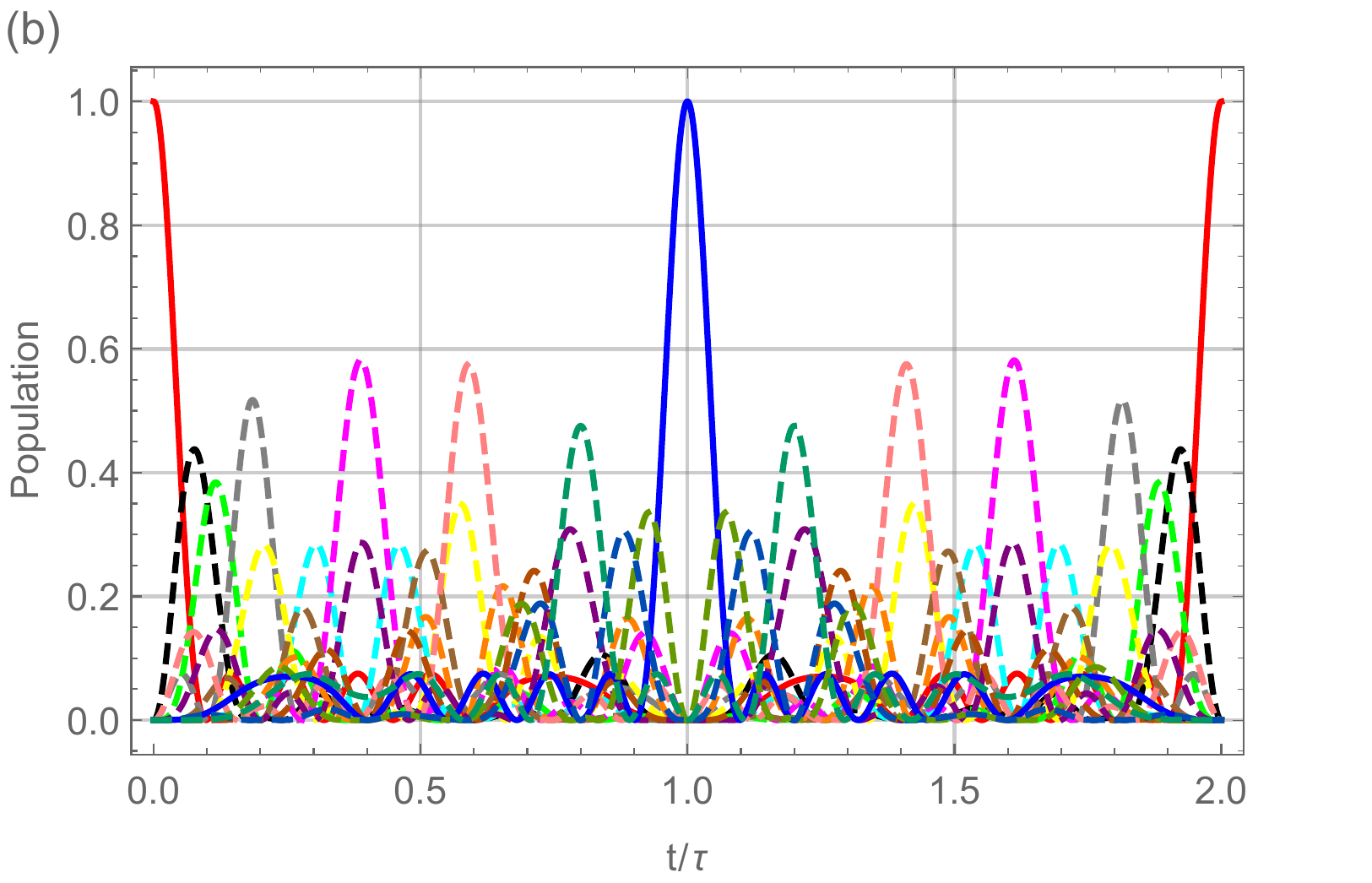}
    \includegraphics[width=3.3cm]{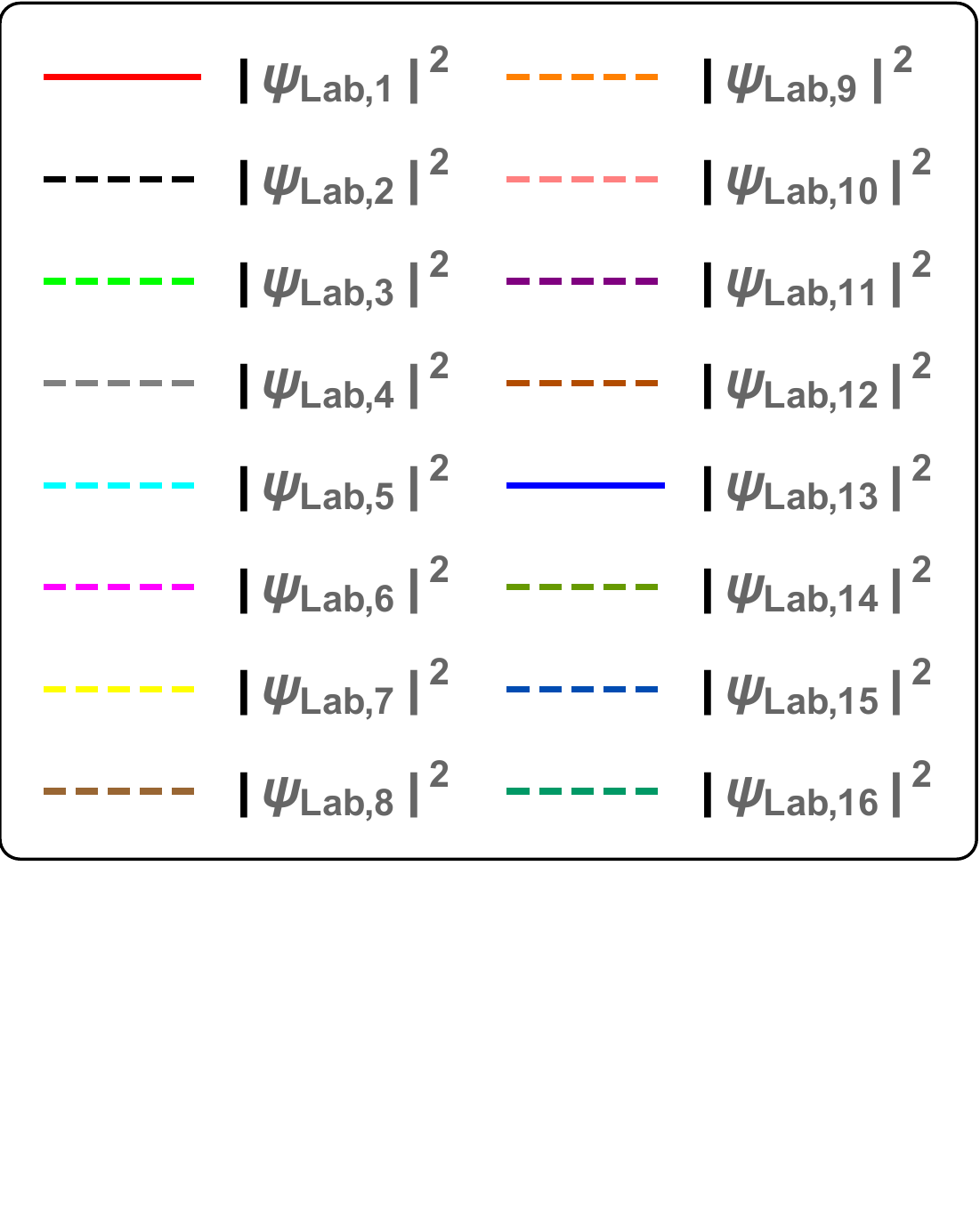}

    \caption{The dynamics of Pythagorean 16-level systems: Numerical simulations of the CPT between $\left|\left.1\right\rangle \right.$ and $\left|\left.13\right\rangle \right.$ in the Lab frame of the 16-level system are shown for two different PPTs: (a) $p=3,q=1,k=0$ $\rightarrow$ $\left(a,b,c\right)=\left(4,3,5\right),k=0$; (b) $p=5,q=1,k=0$ $\rightarrow$ $\left(a,b,c\right)=\left(12,5,13\right),k=0$. The evolution of the population in each state is described as a function of time (measured in $\tau$ units) when the system is prepared in the ground state $\left|\left.1\right\rangle \right.$.}
    \label{fig:numerical_16}
\end{figure*}

The TP frame Hamiltonian and the Lab frame Hamiltonian are both presented in Appendix \ref{appendix:16-level_hamiltonians}. For simplicity, we illustrate the couplings in the Lab frame by an undirected graph in Figure \ref{fig:16_level}. We can see also in Figure \ref{fig:numerical_16} that numerical results confirm the periodic CPT between $\left|\left.1\right\rangle \right.$ and $\left|\left.13\right\rangle \right.$.

The rotation matrix for the case $N=3$, which means that $n=8$ and we work with a 64-level system, is presented in Appendix \ref{appendix:64-level_matrix}.

It is important to realize that the fact that $V\left(I_n\right)$ goes to $V\left(e^{i\pi J^{\left(n\right)}_2}\right)$ in the TP frame through $U=u_1\otimes u_2$ is nothing but a manifestation of the fact that we are working in different representations of the same group. The question to be asked is: Is this a CPT in every representation? In order for it to be a CPT in every representation, the two states must be orthogonal. It is trivial to understand that this holds when $n$ is even, since $e^{i\pi J_y}$ always has alternating 1s and ($\mathrm{-}$1)s on the anti-diagonal, and 0s elsewhere.

So, for every even $n$ we have this CPT from $\frac{1}{\sqrt{n}}V\left(I_n\right)$ to $\frac{1}{\sqrt{n}}V\left(e^{i\pi J^{\left(n\right)}_2}\right)$. Thus, any orthogonal rotation which has the two rows $\frac{1}{\sqrt{n}}V^T\left(I_n\right)$ and $\frac{1}{\sqrt{n}}V^T\left(e^{i\pi J^{\left(n\right)}_2}\right)$ would lead us to a frame in which we have a CPT there (between these two states). What is special in the $n=2^N$ cases is that we can find there a Lab frame in which the Hamiltonian has many symmetries that can be seen via coupling diagrams or coupling undirected graphs, and all of the states are maximally entangled states.

\section{The quantum retrograde canon point of view}
We will now relate the CPT scheme we built to the quantum retrograde canon \cite{retrograde}, investigate the retrograde canon for other unitary groups and then give a more general CPT recipe.

In fact, the CPT we found here in higher dimensions is related to the retrograde canon procedure \cite{retrograde}. We will discuss it from a new group theoretical perspective, which would allow its subsequent generalization to higher unitary groups. Here we give a one-direction claim, and from the reversibility of the proof we deduce that the other direction also holds. If $H\left(t\right)$ is the Hamiltonian of a two level system, and its time evolution operator,
\begin{equation}
    U\left(t,t_0\right)=\mathcal{T}{\mathrm{exp} \left(-\frac{i}{\hslash}\int^t_{t_0}{dt' H\left(t'\right)}\right)} ,
    \label{eqn:propagator}
\end{equation}
satisfies
\begin{equation}
U\left(T,0\right)\left|\left.\uparrow \right\rangle \right.=-\left|\left.\downarrow \right\rangle \right.
\end{equation}
(from unitarity we immediately see that $U\left(T,0\right)=\left|\left.\uparrow \right\rangle \right.\left\langle \left.\downarrow \right|\right.-\left|\left.\downarrow \right\rangle \right.\left\langle \left.\uparrow \right|\right.=\Sigma_2$), then the Hamiltonian
\begin{equation}
    \mathcal{H}\left(t\right)=-H\left(T-t\right)\otimes I+I\otimes H\left(t\right)  ,
    \label{eqn:retrogradeHamiltonian}
\end{equation}
whose propagator is
\begin{equation}
    \mathcal{U}\left(t,t_0\right)=U\left(T-t,T-t_0\right)\otimes U\left(t,t_0\right)  ,
    \label{eqn:retrogradePropagator}
\end{equation}
satisfies
\begin{equation}
\mathcal{U}\left(\frac{T}{2},0\right)V\left(I\right)=V\left( \Sigma_2
\right)  .
\end{equation}

The proof proceeds as follows: Since we know that
$U\left(T,0\right)=\Sigma_2$ and $\Sigma_2$ satisfies $u\Sigma_2u^T=\Sigma_2$ for every $u\in SU\left(2\right)$, we can see that
\begin{equation}
U\left(\frac{T}{2},T\right)U\left(T,0\right)U^T\left(\frac{T}{2},T\right)=\Sigma_2  .
\end{equation}
And since the propagator satisfies $U\left(t_3,t_2\right)U\left(t_2,t_1\right)=U(t_3,t_1)$ (for every $t_1,t_2,t_3$) we get
\begin{equation}
U\left(\frac{T}{2},0\right)U^T\left(\frac{T}{2},T\right)=\Sigma_2  ,
\end{equation}
and therefore (See Appendix \ref{appendix:vectorization}:
\begin{equation}
\begin{split}
\mathcal{U}\left(\frac{T}{2},0\right)V\left(I\right)&=U\left(\frac{T}{2},T\right)\otimes U\left(\frac{T}{2},0\right)V\left(I\right) \\
&=V\left(U\left(\frac{T}{2},0\right)IU^T\left(\frac{T}{2},T\right)\right) \\
&=V\left(\Sigma_2\right)  .
\end{split}
\end{equation}

We can see in the proof that all the steps are reversible, so that the other direction holds. We can summarize:
\begin{equation}
U\left(T,0\right)\left|\left.\uparrow \right\rangle \right.=-\left|\left.\downarrow \right\rangle \right. \Leftrightarrow \mathcal{U}\left(\frac{T}{2},0\right)V\left(I\right)=V\left(\Sigma_2\right) ;
\end{equation}
and the Pythagorean Hamiltonian that we deal with can be obtained from this procedure \cite{retrograde}. Notice that $\Sigma_2=e^{i\pi\frac{\sigma_2}{2}}=e^{i\pi J^{\left(2\right)}_2}$, so we can extend our claim to other representation, by simply replacing $\Sigma_2$ by $Y_n=e^{i\pi J^{\left(n\right)}_2}$.

A natural question arises: Can we claim a similar retrograde statement for the other unitary groups $\mathrm{SU}\left(n\right)$ (when $n>2$)? The answer is no; and the deep reason of this lies in the fact that the only group who is self-dual (the dual of each irreducible representation is isomorphic to it) among these is just $\mathrm{SU}\left(2\right)$.

For every $\mathrm{SU}\left(n\right)$, $n\otimes\overline{n}=1\oplus (n^2-1)$, which means that there is a scalar state which does not change under $U\otimes\overline{U}$, which is of course the state proportional to $V\left(I\right)$, since $\overline{U}=U^{*}$. With this in mind we define first the semi-retrograde Hamiltonian
\begin{equation}
\mathcal{H}\left(t\right)=H^*\left(T-t\right)\otimes I+I\otimes H\left(t\right)
\end{equation}
whose propagator is
\begin{equation}
\mathcal{U}\left(t,t_0\right)=U^*\left(T-t,T-t_0\right)\otimes U\left(t,t_0\right)
\end{equation}
and immediately conclude that
\begin{equation}
U\left(T,0\right)=I \iff \mathcal{U}\left(\frac{T}{2},0\right)V\left(I\right)=V\left(I\right),
\end{equation}
where the proof is very similar to what we did in the retrograde canon's proof. However, this does not give a CPT recipe.

$\mathrm{SU}\left(2\right)$ is special since it is self-dual: Any irreducible representation of $\mathrm{SU}\left(2\right)$ is isomorphic to its dual representation. A representative example would be the simple representation called $2$. $\overline{\mathrm{2}}$ is isomorphic to $\mathrm{2}$ via the transformation $Y\mathrm{=}e^{i\frac{\pi }{\mathrm{2}}{\sigma }_y}$. Only because of this unique feature of self-duality of $\mathrm{SU}\left(2\right)$, we could get our CPT from $V\left(I\right)$ to $V\left(Y\right)$.

In fact, for any $\mathrm{SU}\left(n\right)$, if we have two dual isomorphic representations (of the same dimension) $\rho$ and $\rho^*$, where the isomorphism is the matrix $Y^\dagger$ (it has to be unitary); i.e.
\begin{equation}
    Y^\dagger \rho(g)Y=\rho^*(g) \ \ \forall g \in \mathrm{SU}\left(n\right),
\end{equation}
we get $u Y u^T=Y$ for every $u\in \mathrm{SU}\left(n\right)$, and the same proof scheme holds here too. We get in this case:
\begin{equation}
U\left(T,0\right)=Y \iff \mathcal{U}\left(\frac{T}{2},0\right)V\left(I\right)=V\left(Y\right).
\label{eqn:retro_quaternionic}
\end{equation}
where $U\left(t,t_0\right)$ and $\mathcal{U}\left(t,t_0\right)$ are defined as in Equations (\ref{eqn:propagator}),(\ref{eqn:retrogradePropagator}) and $H\left(t\right)\in\rho$. This is a CPT if and only if $\mathrm{tr}\left(Y\right)=0$; which is equivalent to orthogonality between $V\left(I\right)$ and $V\left(Y\right)$.

Although we could not get an analogous CPT procedure in other special unitary groups, we still can generate a similar argument for a general multi-state Hamiltonian. It is not the same since it does not depend on a singlet state of a group, and it also deals with any quantum mechanical system. The similarity is for the states we use in this argument, and in the case of 2-level systems it reduces to the retrograde canon we presented before. The statement goes as follows: For a general multi-state Hamiltonian $H\left(t\right)$ whose propagator is $U\left(t,t_0\right)$ (Equation (\ref{eqn:propagator})): If there are two normalized states $\left|\left.i\right\rangle \right.$ and $\left|\left.f\right\rangle \right.$ such that:

\begin{enumerate}
\item  $\left|\left\langle i\mathrel{\left|\vphantom{i f}\right.\kern-\nulldelimiterspace}f\right\rangle \right|\mathrm{<1}$ ,

\item  $U\left(T,0\right)\left|\left.i\right\rangle \right.\mathrm{=}\left|\left.f\right\rangle \right.$ , and

\item  $U\left(T,0\right)\left|\left.f\right\rangle \right.\mathrm{=}e^{i\phi }\left|\left.i\right\rangle \right.$, where $\phi $ is real,
\end{enumerate}
then (we define the retrograde Hamiltonian and its propagator as in Equations (\ref{eqn:retrogradeHamiltonian}) and (\ref{eqn:retrogradePropagator}), and define the two states $\left|\left.g\right\rangle \right.\equiv U\left(\frac{T}{\mathrm{2}},0\right)\left|\left.i\right\rangle \right.$ and $\left|\left.h\right\rangle \right.\equiv U\left(\frac{T}{\mathrm{2}},0\right)\left|\left.f\right\rangle \right.$):
\begin{equation}
\begin{split}
\mathcal{U}&\left(\frac{T}{\mathrm{2}},0\right)\left(\mathrm{-}e^{i\phi }\left|\left.ii\right\rangle \right.\mathrm{+}\left|\left.ff\right\rangle \right.\right)\\
&\mathrm{=}U\left(\frac{T}{\mathrm{2}},T\right)\mathrm{\otimes }U\left(\frac{T}{\mathrm{2}},0\right)\left(\mathrm{-}e^{i\phi }\left|\left.ii\right\rangle \right.\mathrm{+}\left|\left.ff\right\rangle \right.\right)\\
&\mathrm{=}U\left(\frac{T}{\mathrm{2}},0\right)\mathrm{\otimes }U\left(\frac{T}{\mathrm{2}},0\right)\left(\mathrm{-}\left|\left.fi\right\rangle \right.\mathrm{+}\left|\left.if\right\rangle \right.\right)\\
&\mathrm{=-}\left|\left.hg\right\rangle \right.\mathrm{+}\left|\left.gh\right\rangle \right.  ,
\end{split}
\end{equation}
and this is a CPT (normalization is needed of course).

If the Hamiltonian $H\left(t\right)$ is time-independent, the conditions reduce to:

\begin{enumerate}
\item  $U\left(2T\right)\left|\left.i\right\rangle \right.\mathrm{=}e^{i\phi }\left|\left.i\right\rangle \right.$, and

\item  $\left|\left\langle i\mathrel{\left|\vphantom{i U\left(T\right) i}\right.\kern-\nulldelimiterspace}U\left(T\right)\mathrel{\left|\vphantom{i U\left(T\right) i}\right.\kern-\nulldelimiterspace}i\right\rangle \right|<1$ ,
\end{enumerate}
and we get CPTs not just from $\mathrm{-}e^{i\phi }\left|\left.ii\right\rangle \right.\mathrm{+}\left|\left.ff\right\rangle \right.$, but also from $U\left(t\right)\otimes U\left(t\right)$ acting on $\mathrm{-}e^{i\phi }\left|\left.ii\right\rangle \right.\mathrm{+}\left|\left.ff\right\rangle \right.$ for every $t$ (as an initial state).
From the argument presented here we can understand that there are basic CPTs in our system, and from them we can build our universal CPT in any even-dimensional representation. We can also investigate the problem with odd-dimensional representations again from another point of view. For more details see Appendix \ref{appendix:basic_CPTs} and Appendix \ref{appendix:the_problem_of_odd}. Moreover, for the time-independent Hamiltonian case, we always can satisfy the two conditions if we start with a combination of two eigen-states (with two nonzero coefficients), and hence they are always fulfilled in two level systems. Other cases of multi-level systems and general time-dependent Hamiltonian may be investigated using the Poincar\'e recurrence theorem \cite{poincare_1,poincare_2,poincare_3,poincare_4}.

\section{Conclusion}
We presented a novel way of manipulating maximally entangled states in $2^{2N}$-level systems using a generalization of the Pythagorean Triples coupling scheme. For this we used a basis of maximally entangled states in which the Hamiltonian has a realistic structure and symmetry of its nonvanishing matrix elements, with a substantially-reduced number of couplings, allowing enormous simplification of its experimental realization in maximally entangled state control. Other $\left(2n\right)^2$-level systems do have the same CPT, but we could not build for them real Lab Hamiltonians that had the same symmetries of the couplings. We found that this scheme, which is based on the quantum retrograde canon, is unique for $\mathrm{SU}\left(2\right)$ and gave a similar argument (which does not give a CPT by itself) for $\mathrm{SU}\left(n\right)$ where $n\ge3$. In addition, we derived a new more general CPT recipe for general multi-state systems.

Other groups and schemes could be considered in a similar way in order to build schemes for them if it is possible, but an immediate case of interest is the quaternionic (or pseudoscalar) representations of $\mathrm{SU}\left(n\right)$ which exist for $n=4k+2$. In this case, the representation is self-dual (the dual representation is isomorphic to to the representation itself), so that it is tempting to check what happens there with the retrograde canon scheme. Another tempting procedure to try to make is to derive a similar scheme for other self-dual groups, for which every irreducible representation is isomorphic to its dual. 

We believe that our analytical schemes and our \emph{natural maximally entangled bases} will offer a new platform for quantum control and quantum information processing of multi-state dynamics.

We would like to thank A. Padan for useful discussions. M.G. gratefully acknowledges support by the Israel Science Foundation (Grant No. 227/15) and the US-Israel Binational Science Foundation (Grant No. 2016224). H.S. gratefully acknowledges support by the Israel Science foundation (Grant No. 1433/15) as well.

\bibliography{apssamp}

\newpage

\appendix

\section{Vectorization function and its inverse}
\label{appendix:vectorization}
Let $X$ be an $m\times n$ matrix. $V\left(X\right)$ simply creates an $mn$-long column vector by stacking $X$'s columns one after the other. We can define this more formally as follows. Let $e_i$ denote the $n\times 1$ matrix (column vector) which is zero everywhere except the $i$-th component, which is 1 (in other words ${\left(e_i\right)}_{j1}={\delta }_{ij}$). Define $E_i$ to be
\begin{equation}
  E_i=e_i\otimes I_m    .
\end{equation}
The vectorization function is then
\begin{equation}
  V\left(X\right)=\sum^n_{i=1}{E_iXe_i}  ,  
\end{equation}
and its inverse is
\begin{equation}
  V^{-1}\left(Y\right)=\sum^n_{i=1}{E^T_iYe^T_i}  .  
\end{equation}
A very useful property of this operator is
\begin{equation}
 V\left(AXB\right)=\left(B^T\otimes A\right)V\left(X\right)  ,   
\end{equation}
and we use it frequently in this paper.

\section{The Hamiltonian of\\the 16-level system in the two frames}
\label{appendix:16-level_hamiltonians}

Here we present explicitly the 16-level system Hamiltonian in the two frames- the TP frame and the Lab frame, as well as the symmetric orthogonal transformation matrix $W$ that takes us from one basis to another.

\begin{widetext}

\tiny{
\begin{equation}
\begin{split}
&H_{TP}=\\
&\left( \begin{array}{cccccccccccccccc}
3V_{14} & \sqrt{3}\Omega_2 & 0 & 0 & \sqrt{3}\Omega_1 & 0 & 0 & 0 & 0 & 0 & 0 & 0 & 0 & 0 & 0 & 0 \\ 
\sqrt{3}\Omega_2 & 3\Delta_1+\Delta_2 & 2\Omega_2 & 0 & 0 & \sqrt{3}\Omega_1 & 0 & 0 & 0 & 0 & 0 & 0 & 0 & 0 & 0 & 0 \\ 
0 & 2\Omega_2 & 3\Delta_1-\Delta_2 & \sqrt{3}\Omega_2 & 0 & 0 & \sqrt{3}\Omega_1 & 0 & 0 & 0 & 0 & 0 & 0 & 0 & 0 & 0 \\ 
0 & 0 & \sqrt{3}\Omega_2 & 3V_{23} & 0 & 0 & 0 & \sqrt{3}\Omega_1 & 0 & 0 & 0 & 0 & 0 & 0 & 0 & 0 \\ 
\sqrt{3}\Omega_1 & 0 & 0 & 0 & \Delta_1+3\Delta_2 & \sqrt{3}\Omega_2 & 0 & 0 & 2\Omega_1 & 0 & 0 & 0 & 0 & 0 & 0 & 0 \\ 
0 & \sqrt{3}\Omega_1 & 0 & 0 & \sqrt{3}\Omega_2 & V_{14} & 2\Omega_2 & 0 & 0 & 2\Omega_1 & 0 & 0 & 0 & 0 & 0 & 0 \\ 
0 & 0 & \sqrt{3}\Omega_1 & 0 & 0 & 2\Omega_2 & V_{23} & \sqrt{3}\Omega_2 & 0 & 0 & 2\Omega_1 & 0 & 0 & 0 & 0 & 0 \\ 
0 & 0 & 0 & \sqrt{3}\Omega_1 & 0 & 0 & \sqrt{3}\Omega_2 & \Delta_1-3\Delta_2 & 0 & 0 & 0 & 2\Omega_1 & 0 & 0 & 0 & 0 \\ 
0 & 0 & 0 & 0 & 2\Omega_1 & 0 & 0 & 0 & 3\Delta_2-\Delta_1 & \sqrt{3}\Omega_2 & 0 & 0 & \sqrt{3}\Omega_1 & 0 & 0 & 0 \\ 
0 & 0 & 0 & 0 & 0 & 2\Omega_1 & 0 & 0 & \sqrt{3}\Omega_2 & -V_{23} & 2\Omega_2 & 0 & 0 & \sqrt{3}\Omega_1 & 0 & 0 \\ 
0 & 0 & 0 & 0 & 0 & 0 & 2\Omega_1 & 0 & 0 & 2\Omega_2 & -V_{14} & \sqrt{3}\Omega_2 & 0 & 0 & \sqrt{3}\Omega_1 & 0 \\ 
0 & 0 & 0 & 0 & 0 & 0 & 0 & 2\Omega_1 & 0 & 0 & \sqrt{3}\Omega_2 & -\Delta_1-3\Delta_2 & 0 & 0 & 0 & \sqrt{3}\Omega_1 \\ 
0 & 0 & 0 & 0 & 0 & 0 & 0 & 0 & \sqrt{3}\Omega_1 & 0 & 0 & 0 & -3V_{23} & \sqrt{3}\Omega_2 & 0 & 0 \\ 
0 & 0 & 0 & 0 & 0 & 0 & 0 & 0 & 0 & \sqrt{3}\Omega_1 & 0 & 0 & \sqrt{3}\Omega_2 & \Delta_2-3\Delta_1 & 2\Omega_2 & 0 \\ 
0 & 0 & 0 & 0 & 0 & 0 & 0 & 0 & 0 & 0 & \sqrt{3}\Omega_1 & 0 & 0 & 2\Omega_2 & -3\Delta_1-\Delta_2 & \sqrt{3}\Omega_2 \\ 
0 & 0 & 0 & 0 & 0 & 0 & 0 & 0 & 0 & 0 & 0 & \sqrt{3}\Omega_1 & 0 & 0 & \sqrt{3}\Omega_2 & -3V_{14} \end{array}
\right)
\end{split}
\end{equation}
}

\normalsize{
\begin{equation}
\begin{split}
W_{16\times 16}=\frac{1}{2}( &\underline{00},\underline{01},\underline{10},\underline{11},\underline{31},\underline{30},\underline{21},\underline{20},\\
&\underline{23},\underline{22},\underline{33},\underline{32},\underline{12},\underline{13},\underline{02},\underline{03} ) \\
&= \frac{1}{2} \left(
\begin{array}{cccccccccccccccc}
1 & 0 & 0 & 0 & 0 & 1 & 0 & 0 & 0 & 0 & 1 & 0 & 0 & 0 & 0 & 1 \\
0 & 1 & 0 & 0 & 1 & 0 & 0 & 0 & 0 & 0 & 0 & 1 & 0 & 0 & 1 & 0 \\
0 & 0 & 1 & 0 & 0 & 0 & 0 & 1 & 1 & 0 & 0 & 0 & 0 & 1 & 0 & 0 \\
0 & 0 & 0 & 1 & 0 & 0 & 1 & 0 & 0 & 1 & 0 & 0 & 1 & 0 & 0 & 0 \\
0 & 1 & 0 & 0 & 1 & 0 & 0 & 0 & 0 & 0 & 0 & -1 & 0 & 0 & -1 & 0 \\
1 & 0 & 0 & 0 & 0 & 1 & 0 & 0 & 0 & 0 & -1 & 0 & 0 & 0 & 0 & -1 \\
0 & 0 & 0 & 1 & 0 & 0 & 1 & 0 & 0 & -1 & 0 & 0 & -1 & 0 & 0 & 0 \\
0 & 0 & 1 & 0 & 0 & 0 & 0 & 1 & -1 & 0 & 0 & 0 & 0 & -1 & 0 & 0 \\
0 & 0 & 1 & 0 & 0 & 0 & 0 & -1 & -1 & 0 & 0 & 0 & 0 & 1 & 0 & 0 \\
0 & 0 & 0 & 1 &0 & 0 & -1 & 0 & 0 & -1 & 0 & 0 & 1 & 0 & 0 & 0 \\
1 & 0 & 0 & 0 & 0 & -1 & 0 & 0 & 0 & 0 & -1 & 0 & 0 & 0 & 0 & 1 \\
0 & 1 & 0 & 0 & -1 & 0 & 0 & 0 & 0 & 0 & 0 & -1 & 0 & 0 &1 & 0 \\
0 & 0 & 0 & 1 & 0 & 0 & -1 & 0 & 0 & 1 & 0 & 0 & -1 & 0 & 0 & 0 \\
0 & 0 & 1 & 0 & 0 & 0 & 0 & -1 & 1 & 0 & 0 & 0 & 0 & -1 & 0 & 0 \\
0 & 1 & 0 & 0 & -1 & 0 & 0 & 0 & 0 & 0 & 0 & 1 & 0 & 0 & -1 & 0 \\
1 & 0 & 0 & 0 & 0 & -1 & 0 & 0 & 0 & 0 & 1 & 0 & 0 & 0 & 0 & -1 \\
\end{array}
\right)
\end{split}
\end{equation}
}

\scriptsize{
\begin{equation}
\begin{split}
H&_{\mathrm{Lab}}=\\
&=
\left( \begin{array}{cccccccccccccccc}
0 & \sqrt{3}V_{12} & 0 & V_{12} & 0 & 2V_{14} & 0 & 0 & 0 & -V_{12} & 0 & 0 & 0 & 0 & 0 & V_{14} \\ 
\sqrt{3}V_{12} & 0 & V_{12} & 0 & 2V_{14} & 0 & 0 & 0 & V_{34} & 0 & 0 & 0 & 0 & 0 & V_{23} & 0 \\ 
0 & V_{12} & 0 & \sqrt{3}V_{12} & 0 & 0 & 0 & 2V_{23} & 0 & 0 & 0 & V_{34} & 0 & V_{14} & 0 & 0 \\ 
V_{12} & 0 & \sqrt{3}V_{12} & 0 & 0 & 0 & 2V_{23} & 0 & 0 & 0 & -V_{12} & 0 & V_{23} & 0 & 0 & 0 \\ 
0 & 2V_{14} & 0 & 0 & 0 & \sqrt{3}V_{12} & 0 & V_{34} & 0 & 0 & 0 & V_{23} & 0 & V_{12} & 0 & 0 \\ 
2V_{14} & 0 & 0 & 0 & \sqrt{3}V_{12} & 0 & V_{34} & 0 & 0 & 0 & V_{14} & 0 & -V_{34} & 0 & 0 & 0 \\ 
0 & 0 & 0 & 2V_{23} & 0 & V_{34} & 0 & \sqrt{3}V_{12} & 0 & V_{23} & 0 & 0 & 0 & 0 & 0 & -V_{34} \\ 
0 & 0 & 2V_{23} & 0 & V_{34} & 0 & \sqrt{3}V_{12} & 0 & V_{14} & 0 & 0 & 0 & 0 & 0 & V_{12} & 0 \\ 
0 & V_{34} & 0 & 0 & 0 & 0 & 0 & V_{14} & 0 & \sqrt{3}V_{34} & 0 & V_{12} & 0 & 2V_{23} & 0 & 0 \\ 
-V_{12} & 0 & 0 & 0 & 0 & 0 & V_{23} & 0 & \sqrt{3}V_{34} & 0 & V_{12} & 0 & 2V_{23} & 0 & 0 & 0 \\ 
0 & 0 & 0 & -V_{12} & 0 & V_{14} & 0 & 0 & 0 & V_{12} & 0 & \sqrt{3}V_{34} & 0 & 0 & 0 & 2V_{14} \\ 
0 & 0 & V_{34} & 0 & V_{23} & 0 & 0 & 0 & V_{12} & 0 & \sqrt{3}V_{34} & 0 & 0 & 0 & 2V_{14} & 0 \\ 
0 & 0 & 0 & V_{23} & 0 & -V_{34} & 0 & 0 & 0 & 2V_{23} & 0 & 0 & 0 & \sqrt{3}V_{34} & 0 & V_{34} \\ 
0 & 0 & V_{14} & 0 & V_{12} & 0 & 0 & 0 & 2V_{23} & 0 & 0 & 0 & \sqrt{3}V_{34} & 0 & V_{34} & 0 \\ 
0 & V_{23} & 0 & 0 & 0 & 0 & 0 & V_{12} & 0 & 0 & 0 & 2V_{14} & 0 & V_{34} & 0 & \sqrt{3}V_{34} \\ 
V_{14} & 0 & 0 & 0 & 0 & 0 & -V_{34} & 0 & 0 & 0 & 2V_{14} & 0 & V_{34} & 0 & \sqrt{3}V_{34} & 0 \end{array}
\right)
  .
\end{split}
\end{equation}
}

\end{widetext}

\section{The transformation in the 64-level system}
\label{appendix:64-level_matrix}

For $N=3$ (i.e., $n=8$, that is, a 64-level system), we only write down explicitly the transformation matrix:
\begin{equation}
\begin{split}
&2\sqrt{2}\ W_{64\times 64}=\\
 & =\begin{array}{cccccccc}
( \underline{000},&\underline{001},&\underline{010},&\underline{011},&\underline{100},&\underline{101},&\underline{110},&\underline{111},\\ 
\underline{031},&\underline{030},&\underline{021},&\underline{020},&\underline{131},&\underline{130},&\underline{121},&\underline{120},\\
\underline{313},&\underline{312},&\underline{303},&\underline{302},&\underline{213},&\underline{212},&\underline{203},&\underline{202},\\
\underline{322},&\underline{323},&\underline{332},&\underline{333},&\underline{222},&\underline{223},&\underline{232},&\underline{233},\\
\underline{230},&\underline{231},&\underline{220},&\underline{221},&\underline{330},&\underline{331},&\underline{320},&\underline{321},\\
\underline{201},&\underline{200},&\underline{211},&\underline{210},&\underline{301},&\underline{300},&\underline{311},&\underline{310},\\
\underline{123},&\underline{122},&\underline{133},&\underline{132},&\underline{023},&\underline{022},&\underline{033},&\underline{032},\\
\underline{112},&\underline{113},&\underline{102},&\underline{103},&\underline{012},&\underline{013},&\underline{002},&\underline{003} )
\end{array}  .
\end{split}
\end{equation}

\section{\label{appendix:basic_CPTs}The more basic CPTs of\\the Pythagorean Hamiltonian}

From the argument presented in Equation (22) in the main text we can understand that there are more basic CPTs in our system, and from them we can build our universal CPT in any even-dimensional representation. Recall that the Hamiltonian
\begin{equation}
H\left(t\right)=\left\{ \begin{array}{ccc}
\Delta_1{\sigma }_z+\Omega_1{\sigma }_x & , & 0\le t<\frac{T}{2} \\ 
-\Delta_2{\sigma }_z-\Omega_2{\sigma }_x & , & \frac{T}{2}\le t\le T \end{array}
\right.
\end{equation}
has a propagator that satisfies
\begin{equation}
 U\left(T,0\right)={\left(-1\right)}^{\frac{p+q}{2}}\left( \begin{array}{cc}
0 & 1 \\ 
-1 & 0 \end{array}  ,
\right)   
\end{equation}
which means that in spin-$\frac{3}{2}$ representation, for example, we have
\begin{equation}
U\left(T,0\right)={\left(-1\right)}^{\frac{p+q}{2}}\left( \begin{array}{cccc}
0 & 0 & 0 & 1 \\ 
0 & 0 & -1 & 0 \\ 
0 & 1 & 0 & 0 \\ 
-1 & 0 & 0 & 0 \end{array}
\right)  .    
\end{equation}
By our claims in the main text, the TP frame's $16\times 16$ Hamiltonian (See Appendix \ref{appendix:16-level_hamiltonians}) fully transfers the state $\frac{1}{\sqrt{2}}\left(e_1+e_{16}\right)$ to another (orthogonal) state at $\frac{T}{2}$, and so does it for the initial state $\frac{1}{\sqrt{2}}\left(e_6+e_{11}\right)$. We will work now with tensor products in order to clarify how we obtain our previous CPT, and we assume, without loss of generality, that ${\left(-1\right)}^{\frac{p+q}{2}}=1$.

According to the CPT scheme we derived in Equation (24) in the main text, the (two independent) `basic' CPTs are:
\begin{subequations}
\begin{equation}
\begin{split}
\frac{1}{\sqrt{2}}&\left(\left|\left.11\right\rangle \right.+\left|\left.44\right\rangle \right.\right)\to\\
&\to \frac{1}{\sqrt{2}}U\left(\frac{T}{\mathrm{2}},0\right)\mathrm{\otimes }U\left(\frac{T}{\mathrm{2}},0\right)\left(\left|\left.41\right\rangle \right.\mathrm{-}\left|\left.14\right\rangle \right.\right)  ,
\end{split}
\end{equation}
\begin{equation}
\begin{split}
\frac{1}{\sqrt{2}}&\left(\left|\left.33\right\rangle \right.+\left|\left.22\right\rangle \right.\right)\to\\
&\to \frac{1}{\sqrt{2}}U\left(\frac{T}{\mathrm{2}},0\right)\mathrm{\otimes }U\left(\frac{T}{\mathrm{2}},0\right)\left(\left|\left.23\right\rangle \right.\mathrm{-}\left|\left.32\right\rangle \right.\right)  ,
\end{split}
\end{equation}
\end{subequations}
and both of them occur at $t=\frac{T}{2}$; and it is easy to see that our known CPT holds at $t=\frac{T}{2}$ too:
\begin{equation}
\begin{split}
&\frac{1}{2}\left(\left|\left.44\right\rangle \right.+\left|\left.33\right\rangle \right.+\left|\left.22\right\rangle \right.+\left|\left.11\right\rangle \right.\right)\to \\
& \to \frac{1}{2}U\left(\frac{T}{\mathrm{2}},0\right)\mathrm{\otimes }U\left(\frac{T}{\mathrm{2}},0\right)\left(\left|\left.41\right\rangle \right.-\left|\left.32\right\rangle \right.+\left|\left.23\right\rangle \right.-\left|\left.14\right\rangle \right.\right)\\
&=\frac{1}{2}\left(\left|\left.41\right\rangle \right.-\left|\left.32\right\rangle \right.+\left|\left.23\right\rangle \right.-\left|\left.14\right\rangle \right.\right)  ,
\end{split}
\end{equation}
where in the last step we used the fact that $uYu^T=Y$ (in every representation). We wrote this last CPT before in another way: $V\left(I_4\right)\to V\left(Y_4\right)$. What makes this CPT unique is that it is universal for every (even) representation, while the `basic' CPTs we have just seen are not. It is easy to see that for every $\left|\left.{\psi }_0\left(\alpha ,\beta \right)\right\rangle \right.=\frac{\alpha }{\sqrt{2}}\left(\left|\left.11\right\rangle \right.+\left|\left.44\right\rangle \right.\right)+\frac{\beta }{\sqrt{2}}\left(\left|\left.33\right\rangle \right.+\left|\left.22\right\rangle \right.\right)$ we have a CPT (${\left|\alpha \right|}^2+{\left|\beta \right|}^2=1$), since the basic CPT's initial and final states are built of different (and orthogonal) one-particle states, so that the final state (at $t=\frac{T}{2}$) we get from propagating $\left|\left.{\psi }_0\left(\alpha ,\beta \right)\right\rangle \right.$ is orthogonal, by definition, to the initial state. But as we have mentioned, with $\alpha =\beta =\frac{1}{2}$ we have a universal CPT.

For a general even representation $n$, we have $\frac{n}{2}$ basic CPTs of the form: ($i\in \left\{1,2,\dots ,\frac{n}{2}\right\}$)
\begin{equation}
\begin{split}
&\frac{1}{\sqrt{2}}\left(\left|\left.ii\right\rangle \right.+\left|\left.n+1-i,n+1-i\right\rangle \right.\right)\to\\
&\to {\left(-1\right)}^i\frac{1}{\sqrt{2}}U\left(\frac{T}{\mathrm{2}},0\right)\mathrm{\otimes }U\left(\frac{T}{\mathrm{2}},0\right)\\
&\ \ \ \ \ \ \ \ \ \ \ \ \ \ \ \ \ \ \ \ \ \ \ \left(\left|\left.i,n+1-i\right\rangle \right.\mathrm{-}\left|\left.n+1-i,i\right\rangle \right.\right) \  ,
\end{split}
\end{equation}
which are to be calculated in that representation. However, as we have discussed, the sum of all these states exhibits a known CPT, which is independent of the parameters $p,q,k$:
\begin{equation}
\frac{1}{\sqrt{n}}\sum^n_{i=1}{\left|\left.ii\right\rangle \right.}\to \frac{1}{\sqrt{n}}\sum^n_{i=1}{{\left(-1\right)}^i\left|\left.i,n+1-i\right\rangle \right.}  ,
\end{equation}
which is absolutely the known one.

\section{\label{appendix:the_problem_of_odd}The issue in odd-dimensional representations}

After Equation (24) in the main text and Section V of this supplemental material, one can clearly see the issue in odd-dimensional representations we discussed in the main text from another point of view. For simplicity we will demonstrate the problem in the 3-level system of spin-1 representation. Assuming that $U\left(T,0\right) = Y$, we get in the spin-1 representation $U\left(T,0\right)=\left( \begin{array}{ccc}
0 & 0 & 1 \\ 
0 & -1 & 0 \\ 
1 & 0 & 0 \end{array}
\right)$, which means:
\begin{subequations}
\begin{equation}
U\left(T,0\right)\left|\left.1\right\rangle \right.\mathrm{=}\left|\left.3\right\rangle \right.  ,
\end{equation}
\begin{equation}
U\left(T,0\right)\left|\left.2\right\rangle \right.\mathrm{=-}\left|\left.2\right\rangle \right.  ,
\end{equation}
\begin{equation}
U\left(T,0\right)\left|\left.3\right\rangle \right.\mathrm{=}\left|\left.1\right\rangle \right.  .
\end{equation}
\end{subequations}
According to Equation (24) in the main text, this guarantees just one CPT in the retrograde canon's TP frame Hamiltonian, $\mathcal{H}\left(t\right)=-H\left(T-t\right)\otimes I+I\otimes H\left(t\right)$:
\begin{equation}
    \begin{split}
        \frac{1}{\sqrt{2}}&\left(\mathrm{-}\left|\left.11\right\rangle \right.\mathrm{+}\left|\left.33\right\rangle \right.\right)\mathrm{\to }\\
        & \mathrm{\to } \frac{1}{\sqrt{2}}U\left(\frac{T}{\mathrm{2}},0\right)\mathrm{\otimes }U\left(\frac{T}{\mathrm{2}},0\right)\left(\mathrm{-}\left|\left.31\right\rangle \right.\mathrm{+}\left|\left.13\right\rangle \right.\right)\\
        &\ \ \ \ \ \ \ \ \ \ \ \ \ \ \ \ \ \ \ \ \ \ \ \ \ \ \ \ \ \ \ \ \ \ \ \ \ \ \ \ \ \ \ at\ \ t=\frac{T}{2} \ ,
    \end{split}
\end{equation}
which is obviously not universal and does depend on more details of $H\left(t\right)$; just like the other analogous `basic' CPTs we have seen in even representations. So we do not have a similar picture that allows us to repeat the same way to build a universal CPT.  It does hold that if we start with $V\left(I\right)$ we end in $V\left(Y\right)$ at $t=\frac{T}{2}$, but we cannot see this directly from Equation (24) in the main text as we did in even representations, and, in any case, this is not a CPT since the two states are not orthogonal.

\end{document}